\documentclass[aps,10pt,bibnotes,prb,amsmath,citeautoscript,twocolumn,
               preprintnumbers,amssymb,amsmath,
               letterpaper]{revtex4-1}

\usepackage{graphicx}
\begin{document}

\title{Interplay between strain and oxygen vacancies in lanthanum aluminate}
\author{Joshua D. Sayre}
\author{Kris T. Delaney}
\affiliation{Materials Department, University of California, Santa Barbara, California 93106, USA}
\author{Nicola A. Spaldin}
\affiliation{Materials Theory, ETH Zurich, Wolfgang-Pauli-Strasse 27, 8093 Z\"{u}rich, Switzerland}

\date{\today}

\begin{abstract}
We evaluate the interplay between epitaxial strain and oxygen vacancy formation in the perovskite-structure 
oxide lanthanum aluminate, LaAlO$_3$. 
Using density functional theory within the GGA$+U$ approximation we calculate the dependence of the oxygen 
vacancy formation energy on the biaxial strain conditions.
We find that the change in formation energy with strain is negligible over the range
of strain values usually accessible through coherent epitaxial growth.
Our findings suggest that experimental reports of different oxygen vacancy concentrations in
strained films result from extrinsic factors, or from other impurities such as defect complexes.
\end{abstract}

\maketitle

\section{Introduction}
Perovskite-structure oxides, ABO$_3$, are of fundamental and potentially technological interest 
because of their numerous and often coupled functionalities, such as ferroelectricity, magnetism, 
multiferroicity, metal-insulator transitions, etc. In recent years, it has been shown that
functional behavior in such oxides can be enabled or enhanced when they are prepared in thin-film
form. For example, ferroelectricity has been reported in thin films of SrTiO$_3$, which is a 
quantum paraelectric in its bulk form, and enhanced polarization has been observed in thin 
films of ferroelectric BaTiO$_3$ \cite{Haeni2004,Choi2004}. In both cases, the improved behavior 
is attributed to the bi-axial epitaxial strain introduced when the film is grown coherently on 
a substrate with a different lattice constant.

It is usually assumed that the change in in-plane lattice constant associated with epitaxial
strain is accommodated through structural distortions such as changing the internal bond lengths,
or changing the angle or pattern of rotations and tilts of the oxygen octahedra 
\cite{Megaw/Darlington_1975,Thomas1996,Rondinelli/Spaldin:2011}. 
An additional possibility is that the film changes its defect profile, for example through
changing the concentration of oxygen vacancies or the degree of cation non-stoichiometry. 
In particular the presence of oxygen vacancies in perovskite oxides is believed to result in 
an expansion of the crystal lattice due to underbonding caused by the two missing oxygen
electrons \cite{Adler2001,Ullmann2001}. This suggests that the increase in volume associated
with biaxial tensile strain might also increase the concentration of oxygen vacancies.  
Such changes are of profound importance as the presence of anti-site defects, interstitials,
vacancies and/or defect complexes can have striking effects on the properties of perovskite
oxides. 

Any correlation between epitaxial strain and intrinsic defect profile is difficult to 
quantify experimentally. Strain is introduced by changing the substrate, which necessarily
leads to other changes in growth conditions; the different growth conditions can then cause
changes in the concentration and type of defects that might be unrelated to strain.
In this work we use first-principles density functional theory to investigate the relationship
between strain and the formation energy of oxygen vacancies in the model perovskite-structure
oxide, lanthanum aluminate, LaAlO$_3$ (LAO).
Bulk, strain-free LAO has $R\bar{3}c$ symmetry with a slight distortion of the rhombohedral angle
away from the ideal 60$^{\circ}$. The oxygen octahedra rotate by $\sim$6.3$^{\circ}$ in alternating 
directions around the pseudocubic [111] direction, in the a$^-$a$^-$a$^-$ Glazer notation tilt
pattern \cite{Lehnert2000,Glazer1972,Muller1968}. It has a large band gap of ∼5.6 eV \cite{Lim2002}, and so is robustly insulating even within the local density or generalized gradient approximations to density functional theory, which are known to systematically underestimate electronic band gaps \cite{Perdew1983}. This is particularly important for first-principles studies of defective systems, so that defect-related transition levels which are experimentally found to be in the band gap do not erroneously fall outside of the band gap, changing the predicted electrical characteristics. In addition, LAO is structurally robust with no tendency to Jahn-Teller or ferroelectric distortion.  

First-principles calculations were used recently to determine the intrinsic structural distortions 
induced by changes in the in-plane lattice constants of LaAlO$_3$ \cite{Hatt2010}. It was found that 
the [111]-oriented
rotation pattern of bulk LAO is stable only for a small in-plane biaxial strain region (-0.2 to 0.1 \%)
and larger strains result in different patterns. Biaxial compression 
changes the rotation scheme to a$^0$a$^0$c$^-$, in which the octahedra rotate alternately 
along the [001] direction, and tilts around the in-plane axes are suppressed. For biaxial 
tension, the octahedral rotation scheme changes to a$^-$a$^-$c$^0$, with alternating
rotations around the [110] direction. 

First-principles calculations have also been used to explore the structure and formation 
energies of various defects in the high temperature, high symmetry cubic $Pm\bar{3}m$ 
structure of LAO \cite{Xiong2006,Luo2009}.  
Luo et al. calculated the atomic structures, energy levels, and formation energies of charged 
interstitials, vacancies, Frenkel pairs, antisite defects, and Schottky defects. As expected, 
they found that the formation energies of O vacancies are lowest in reducing conditions, 
whereas the La vacancy formation energy decreases with increasing O chemical potential. 
They also found that considerable structural distortion occurred around many of the defects, 
with substantial changes in the inter-ionic distances. It is unclear from Ref.~\onlinecite{Luo2009} whether the
lattice parameters were fixed to the ideal values during the calculations; changes in 
cell parameters associated with the different defect types were not reported.
Likewise the focus of Xiong et al. in Ref.~\onlinecite{Xiong2006} was on the electronic energy 
levels generated by the formation of defects, rather than on any associated changes in lattice parameter.

%
%

Here we present results of our first-principles study combining {\it both} epitaxial strain and 
defects in our calculations. Our goal is to determine whether a critical value of strain
exists at which it is intrinsically more energetically favorable to incorporate 
defects into the sample, instead of further deforming the bond lengths and angles of
the ideal stoichiometric structure.

\section{Computational Details}

Our calculations were performed using density functional theory as implemented in the 
Vienna ab initio simulation package (VASP) \cite{KresseG1996}. The projector augmented 
wave (PAW) method was used for core-valence separation with the supplied VASP potentials
for the La, Al and O atoms \cite{Blochl1994}. Test calculations for bulk LaAlO$_3$ were 
performed using a 10 atom rhombohedral cell (pseudocubic angle fixed at 90 degrees) with a 4x4x4 Monkhorst-Pack $k$-point mesh 
and a plane-wave basis energy cutoff of 600 eV. We defined biaxial strain as  
\begin{equation}
\epsilon = \frac{a-a_0}{a_0}.
\end{equation}
where $a_0$ is the calculated equilibrium lattice constant. Once strained, the out-of plane ($c$) 
lattice parameter was relaxed and the ionic positions optimized until
the forces were below 1meV/\AA.

\subsection{Choice of exchange-correlation functional}
We first tested the behavior of ideal bulk LaAlO$_3$ to select an appropriate exchange-correlation
functional that adequately describes both the structural and electronic properties. 
While the local density approximation (LDA) has been previously shown to give a good description
of the structural properties of LAO \cite{Hatt2010}, it incorrectly places the 
highly localized La \emph{f} states at the bottom of the conduction band \cite{Czyzyk1994}, where they will
likely spuriously interact with oxygen vacancy states in our defect calculations.  
One apparent solution is to apply a Hubbard $U$ correction within the LDA$+U$ formalism; a $U$
value of 11 eV was used previously to correct the La \emph{f} state position in studies for 
which detailed structural distortions were not important \cite{Okamoto2006}. We found,
however, that, while the LDA obtains almost perfect agreement with experiment for the magnitude
of the octahedral rotations, the addition of a $U$ to the La $f$ states strongly suppresses
them giving unacceptable structural properties. In contrast, the 
generalized gradient approximation (GGA), over-estimates the octahedral rotations 
and addition of a $U$ to the La $f$ states reduces them towards the experimental values, while
also correcting the energy of the La $f$ states (Fig. \ref{GGA+U}). Our tests
of electronic properties and structural properties as a function of $U$ led us to a choice of 
$U=10.32$ eV on the La $f$ orbitals as the optimal choice, within PBE GGA \cite{Perdew1996} and
the Dudarev method \cite{Dudarev1998}. Double-counting corrections were made in the fully localized limit.

\begin{figure}
\centering
\includegraphics[scale=0.35]{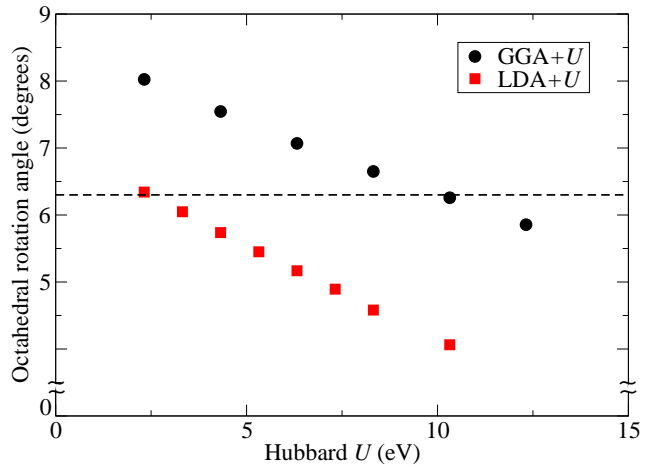}
\caption{Calculated oxygen octahedral rotation angles within the LDA$+U$ and GGA$+U$ methods as a
function of the Hubbard $U$ parameter. The experimental value is shown by the horizontal dashed
line. } 
\label{fbandvsu}
\end{figure}

\begin{figure}
\centering
\includegraphics[scale=0.35]{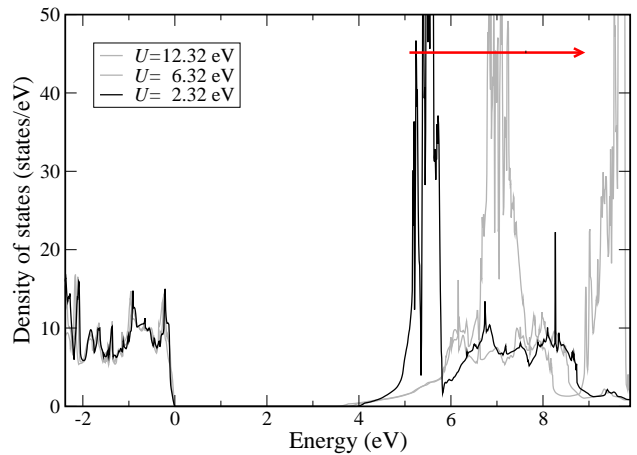}
\caption
{Calculated GGA$+U$ density of states for LaAlO$_3$ for different $U$. In all cases, the top
of the valence band is set to 0eV. As $U$ is increased, 
the La \emph{f} states increase in energy and shift away from the conduction band edge (red
arrow). The valence band density of states is effectively independent of $U$.}
\label{GGA+U}
\end{figure}

\subsection{Oxygen vacancies}

\begin{figure}
\centering
\includegraphics[scale=0.27]{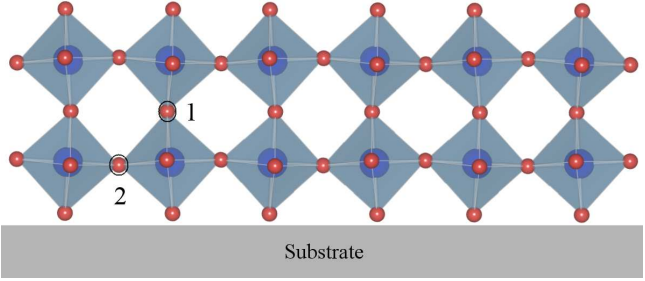}
\caption[Oxygen vacancy location]%
{The two symmetry-inequivalent positions for an oxygen vacancy in biaxially strained
LaAlO$_3$ within
our supercell. In this study we use position 1, in which the oxygen vacancy lies in the normal 
axis to the strain rather than in the basal plane.}
\label{Ovacloc}
\end{figure}

Oxygen vacancy concentrations of $\sim$4\% (one out of 24 atoms) were studied using supercells 
containing 40 atoms (a 2x2x2 supercell of the 10-atom rhombohedral unit cell with the pseudocubic 
angle fixed to 90 degrees); this allowed us to reduce our Monkhorst-Pack mesh to 1x1x1. 
When biaxial strain is included, two distinct placements of the oxygen vacancy -- within the biaxially constrained direction or in the unconstrained relaxed $c$ direction.
(Fig.~\ref{Ovacloc}) -- are possible within our choice of supercell; here we studied the perpendicular 
configuration. 

The oxygen formation energy was also calculated for a larger 4x4x4 supercell to assess proximity
to the
dilute limit. The change in formation energy between the 2x2x2 cell and the 
4x4x4 cell (corresponding to a change in oxygen vacancy concentration from $\sim$4\% to 
$\sim$0.5) was approximately 10meV. 

\section{Results}

\subsection{Calculated structure of unstrained LaAlO$_3$ with oxygen vacancies}

First we introduce an oxygen vacancy into the 40 atom supercell and fully relax the atomic positions with
no strain applied, constraining the lattice parameters to their bulk values. We find that the octahedral rotation pattern changes from the a$^-$a$^-$a$^-$ pattern of bulk LaAlO$_3$ to the a$^-$a$^-$c$^0$, in which the octahedral rotations around the $c$ axis are lost. Interestingly this is the same pattern as found in defect-free LAO under biaxial tension. The deviation from the bulk rotation pattern with the introduction of the oxygen vacancy is likely 
related to the long range coherence of octahedral rotational instabilities, and their interruption
by the presence of the vacancy. This finding that an oxygen vacancy has long range structural implications opens the possibility for strain to change the formation energy for oxygen vacancies.  

\subsection{Electronic properties with oxygen vacancies}

\begin{figure}
\centering
\includegraphics[scale=0.31]{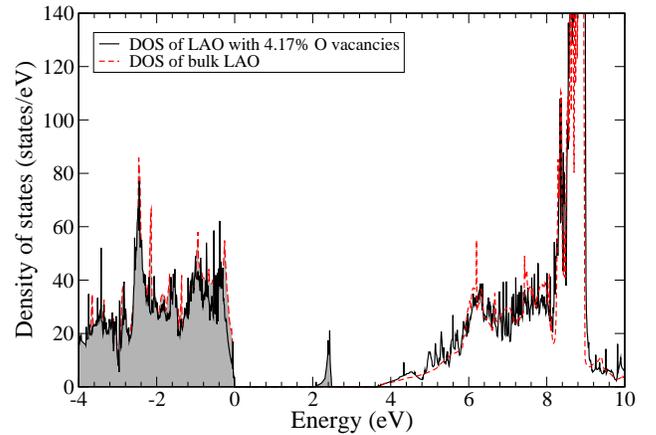}
\caption[Density of states with and without an oxygen vacancy]%
{Calculated densities of states for unstrained LAO with (black line) and without (red line) oxygen vacancies. 
The densities of states are aligned so that the tops of the bulk valence bands lie at 0 eV. The occupied states are shown with shading; in the vacancy structure, the impurity states are fully occupied and lie within
the band gap, ~2.6 eV from the valence band edge.}
\label{DOS}
\end{figure}

Our calculated electronic densities of states for unstrained LAO in both ideal and defective structures
are shown in Figure~\ref{DOS}. The oxygen vacancy state is fully occupied and falls within the band gap of the ideal bulk material. Apart from this isolated impurity band, the ideal and defective densities of states are very similar.  The different k-point sampling between the two calculations causes the slight change in the smoothness of the plots.

\subsection{Combined effect of strain and oxygen vacancies on structure}

We find that
the introduction of an oxygen vacancy into the strained structures does not qualitatively change
the octahedral tilt patterns from those of the corrseponding strained structures without defects.
Under biaxial compression, the defect structure 
adopts the a$^0$a$^0$c$^-$ tilt pattern found in the defect-free structure under biaxial compression;
under biaxial tension the a$^-$a$^-$c$^0$ structure of the unstrained oxygen vacancy structure,
and the defect-free tensile-strained structure is retained. 

\subsection{Strain dependence of oxygen vacancy formation energy}

Next we calculate the formation energy of oxygen vacancies as a function of biaxial strain. 
The formation energy, $\Omega$, of an oxygen vacancy can be calculated from
\begin{equation}
\Omega = E_{tot}(V_{O}^{0})-E_{tot}(bulk)+N\mu_{O}
\end{equation}
where $E_{tot}(V_{O}^{0})$ is the energy of the supercell with an oxygen vacancy, $E_{tot}(bulk)$ 
is the energy of the defect-free supercell, $N$ is the number of oxygen vacancies (one in our case), 
and $\mu_{O}$ is the chemical potential of the oxygen reservoir \cite{VandeWalle2004} . 

The oxygen chemical potential ($\mu_{O}$) is defined as 
\begin{equation}
\mu_{O} = \mu_{O,bulk}-\frac{1}{2}E(O_{2})
\end{equation}
where $\mu_{O,bulk}$ is the bulk chemical potential and $E(O_{2})$ is the formation energy of 
an oxygen molecule. It is clear that the formation energy depends linearly on the oxygen chemical 
potential; at low oxygen over-pressure, the formation energy is lower yielding a higher concentration 
of oxygen vacancies and vice versa. 

\begin{figure}
\centering
\includegraphics[width=0.9\columnwidth]{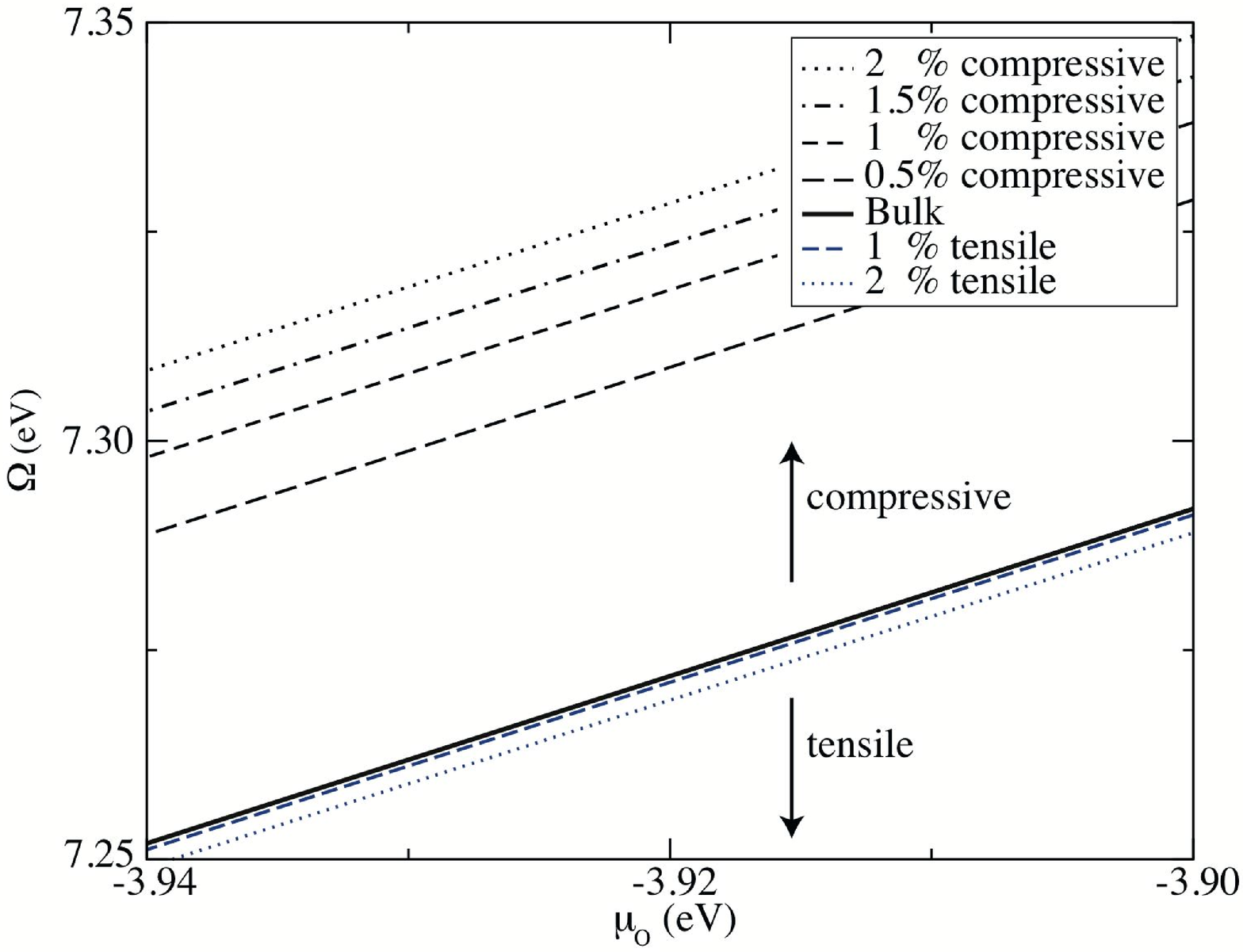}
\caption[Oxygen vacancy formation energy versus chemical potential]%
{Calculated oxygen vacancy formation energies, $\Omega$, as a function of oxygen chemical potential, $\mu_0$
for a range of strain states. Lower chemical
potentials correspond to oxygen-poor conditions and vice versa. Results for four compressive strain  
and four tensile strain values ($\pm$ 0.5\%, 1\%, 1.5\%, and 2\%) are shown. The bulk (unstrained value) falls lies with the tensile strain values. }
\label{Form_Strain_Dep}
\end{figure}

In Fig.~\ref{Form_Strain_Dep} we show our calculated formation energies at different biaxial 
strain values (from -2\% to +2\%) as a function of the oxygen chemical potential. Note that
the formation energies are large -- $>7$eV -- consistent with previous results for related
oxides \cite{Xiong2006,Luo2009}. As expected from our earlier simple volume arguments, the
formation energies at a given chemical potential are higher under biaxial compression than
under tension. The strain dependence of the formation energies is small in comparison to the 
total formation energies, with a 2\% increase in compressive strain increasing the formation 
energies by only $\sim$20 meV. 

We find that the strain dependence of the formation energy at a specific oxygen chemical potential 
is much stronger for compressive than for tensile strains. This is likely a consequence of our
choice of location of the oxygen vacancy within the unit cell: 
Under compression rotations around the [001] direction dominate the structure, and these are 
more strongly affected by the in-plane oxygen vacancy.

\section{Discussion}
In summary, we have shown that density functional theory within the GGA$+U$ approximation
gives a good description of both the structural and electronic properties of LaAlO$_3$, when
a $U$ of 10.32 eV is applied to the La $4f$ states. Perhaps surprisingly, introduction of 
oxygen vacancies at 4\% 
concentration in axial positions induces a change in the tilt pattern from the bulk 
a$^-$a$^-$a$^-$ to the a$^-$a$^-$c$^0$ pattern; this is the pattern that was previously 
identified for defect-free LaAlO$_3$ under tensile strain \cite{Hatt2010}.  

We found that the formation energy for an oxygen vacancy in LaAlO$_3$ is between 3-10 eV 
depending on the oxygen chemical potential. The change in this value with experimentally
accessible strain values are orders of magnitude smaller, with a  $\sim$75 meV increase 
for 2\% compressive strain and 5 meV decrease for 2\% tensile strain. Interestingly, the 
change in vacancy formation energy
with strain is comparable to the change in intrinsic total energy (Fig. 2 of Ref.
\onlinecite{Hatt2010}), suggesting that under suitable conditions strain could be partially
accommodated by change in vacancy concentration. It is also likely, however, that 
strain will modify barriers for defect migration, making a direct experimental comparison
of equilibrium defect concentrations at different strain values fundamentally prohibitive. 

Finally we mention that in perovskite-structure (and other) oxides in which the
B site is a transition metal, the introduction of oxygen vacancies causes a change
in oxidation state of the B site. Such changes in oxidation states are accompanied
by large changes in coordination volume. As a result, stronger strain-dependence
of oxygen vacancy formation energies should be expected. 

\section{Acknowledgements}
This work was supported by NSF DMR-0940420. 
We used the TeraGrid computing facilities at NCSA, and the California
Nanosystems Institute facilities provided by NSF CHE-0321368 and Hewlett-Packard.
NAS was supported by ETH Z\"urich.

\end{document}